\documentclass[12pt]{article}
\usepackage{graphics,graphicx}
\usepackage{times,amsfonts,amssymb,latexsym,amsmath}
\usepackage{url}
\usepackage[T1]{fontenc}              
\usepackage[latin9]{inputenc}         

\usepackage[a4paper]{geometry}        
\usepackage{amsmath,amssymb}          
\usepackage{bm}                       

\usepackage{url}                      
\usepackage{xspace}                   

\usepackage{xspace, setspace}  

\usepackage{graphicx}

\usepackage[square,numbers]{natbib}
\bmdefine\X{X}
\bmdefine\Z{Z}
\bmdefine\x{x}
\bmdefine\z{z}
\bmdefine\y{y}
\bmdefine\Y{Y}

\bmdefine\M{M}
\bmdefine\Q{Q}
\bmdefine\P{P}
\bmdefine\w{w}
\bmdefine\W{W}
\bmdefine\p{p}
\bmdefine\T{T}
\bmdefine\t{t}
\bmdefine\B{B}
\bmdefine\I{I}
\bmdefine\u{u}
\bmdefine\p{p}
\bmdefine\Sig{\Sigma}
\bmdefine\E{E}
\bmdefine\F{F}
\bmdefine\S{S}
\bmdefine\a{a}
\bmdefine\A{A}
\usepackage{bm}

\title{\Large A plea for neutral comparison studies\\ in computational sciences}

\author{Anne-Laure Boulesteix$^{1}$ and Manuel J. A. Eugster$^{2}$}

\begin{document}

\maketitle
\noindent
$^{1}$Department of Medical Informatics, Biometry and Epidemiology, University of Munich, Marchioninistr. 15, 81377 Munich, Germany.\\
$^{2}$Department of Statistics, University of Munich, Ludwigstr. 33, 80539 Munich, Germany.

\subsection*{\centerline{Abstract}}
In a context where most published articles are devoted to the development of \lq\lq new methods'', comparison studies are generally appreciated by readers but surprisingly given poor consideration by many scientific journals. This letter stresses the importance of neutral comparison studies for the objective evaluation of existing methods and the establishment of standards by drawing parallels with clinical research.

\section{Introduction}

The main goal of methodological research in computational sciences
(including, e.g. bioinformatics, machine learning, or computational statistics) is the development of new methods. By
development of new methods, we mean that the researchers suggest new
procedures for analyzing data sets. The new procedure should be
applicable to specific substantive research questions, but these
substantive research questions (often) are not the primary
center of interest of the methodological researcher. A methodological
researcher develops new methods, in contrast to substantive
researchers who apply the methods developed by others, e.g. to their
e.g. genetic or transcriptomic data. New methods are expected to
\lq\lq make the world better'' by, roughly speaking, making the
results of statistical analyses closer to the
truth. Surprisingly, comparison studies and reviews 
  investigating the closeness to the truth are often considered as
less exciting and less useful by many researchers or by most journal
editors, and excluded from the journals' scopes.

This is in strong contrast to clinical research. The ultimate goal in
clinical research is to \lq\lq make the world better'' by somehow
improving the health outcome of patients (or reducing the cost while
maintaining the same outcome), e.g. through a specific drug, therapy
or prevention strategy. Roughly speaking, the clinical analogue of a
computational article suggesting a \lq\lq new method'' would be an
article suggesting a new intervention for improving health
outcome. Yet, most published medical papers do not directly suggest
such a new measure. Many other types of clinical research projects are
conducted, for instance large validation studies, phase IV clinical
trials, or meta-analyses. Of course, crucial differences between
computational science research and medical research make comparisons only
partially pertinent. Research on algorithms and methods does not
follow the same rules as research involving human beings with direct
potentially  vital consequences. The development of a new drug or new
prevention strategy essentially requires more time, money,
coordination and caution than the development of a new statistical
method. Some principles, however, hold for both worlds. If we focus on
the problem of comparison studies considered in this paper, the
question is \lq\lq can we imagine a world in which clinical journals
accept to publish only underpowered phase I or II clinical trials
evaluating new therapies but no phase III or IV trials?'' The answer is
of course no. In  data analysis journals,
however, the equivalent of phase III and IV trials,
i.e. well-conducted comparison studies in our metaphor, are often
considered as not deserving publication.

We claim that comparison studies in computational sciences may however be necessary to ensure that previously proposed methods work as expected in various
situations and that emerging \lq\lq standard practice rules'' adopted
by substantive researchers or statistical consultants are the result
of well-designed studies performed by computational science
experts. The community tends to establish standards and guidelines as
time goes by. In an ideal world, these standards are the results of
well-done comparative studies and consensus from independent
teams. However, other factors might contribute to promote a particular
method, including the reputation of the authors or the impact factor of the
journals the method was published in. From the point of view of
applicants (say, biologists), further criteria
include the availability of well-documented and user-friendly
implementations of the method or an application of this method in one
of the few leading scientific journals that other scientists tend to
imitate. These criteria may seem natural. After all, a method
published by a renown author in an excellent journal is more likely to
work well than a method published by an unknown author in a
low-ranking journal. Availability of good software is of course a
crucial advantage for applicants who would not be able or would not
have time to implement any of the methods themselves. And a method
that worked well in a previous well-published study is perhaps more
likely to also work well in future studies than another method.

It is unclear, however, whether standard practice rules should be
established solely on such \lq\lq subjective'' criteria. Would it
not be better to give more importance to comparison studies? One may
of course argue that comparison studies can be performed within
original articles presenting new methods. Indeed, in practice new
methods are usually compared to a few existing methods in order to
establish their superiority. Such comparison studies are extremely
important for illustrative purposes, i.e. to demonstrate that the
developed method is applicable in practice and yields acceptable
results, but should strictly speaking not be considered as comparison
studies because they are often substantially biased and thus not
\lq\lq neutral''.

For example, in the context of clinical outcome prediction or
diagnosis based on high-dimensional \lq\lq omics data'' (such as,
e.g. microarray gene expression data),
hundreds of articles presenting new supervised classification
algorithms have been published in the 
bioinformatics, statistics and machine learning literature. Almost all
of them claim that the new method \lq\lq performs better'' than
existing methods. Most often these claims are based on small real data
studies including a few exemplary data sets. The fact that for twelve
years hundreds of authors have been claiming that
their new method for classification using microarray data performs
better than existing ones 
  suggests that something goes wrong
in the comparison studies performed in these articles. Similar
discussions can be found in other fields of application of
machine learning and computational statistics
\citep[e.g.,][]{Wagstaff@2012}. What goes wrong? How should a
proper comparison study look like? Is it possible to perform such
comparison studies in the context of an original article presenting a
new method?

\section{Over-optimism and the need for neutral comparison studies}

Comparison studies included in original research articles presenting
new methods are often over-optimistic with respect to the superiority
of the new method. Some reasons for over-optimism have been empirically
assessed   and discussed  in the context of supervised classification using high-dimensional molecular data \cite{boulesteix2010over,jelizarow2010over}. The first and perhaps most
obvious reason for over-optimism is that researchers sometimes
``randomly search'' for a specific data set such that
their new method works better than existing approaches, yielding a
so-called \lq\lq data set bias'' \cite{yousefi2010reporting}. A second source of
over-optimism, which is related to the optimal choice of the data set
mentioned above, is the optimal choice of a particular setting in
which the superiority of the new algorithm is more pronounced. For
example, researchers could report the results obtained after a
particular feature filtering which favors the new algorithm compared
to existing benchmark approaches. The third and probably most subtle
problem is that researchers often tend to optimize their new
algorithms to the data sets they consider during the development phase
\citep{boulesteix2010over,jelizarow2010over}. This mechanism
essentially affects all research fields related to data analysis such
as statistics, machine learning, or bioinformatics. Indeed, the
trial-and-error process constitutes an important component of data
analysis research. As most inventive ideas have to be improved
sequentially before reaching an acceptable maturity, the development
of a new method is per se an unpredictable search process. The problem
is that this search process leads to an artificial optimization of the
method's characteristics to the considered data sets. Hence, the
superiority of the novel method over an existing method (as measured,
e.g. through the difference between the cross-validation error rates)
is sometimes considerably overestimated.

Other reasons are of technical nature and related to the ability of
the researchers to use the compared methods properly. For example, if
an implementation problem occurs with the competing approaches and
slightly worsens their results, researchers often tend to
spontaneously accept these inferior results. Conversely, they would
probably obstinately look for the programming error if such problems
occur with their new algorithm. In the same vein, they may
unintentionally set the parameters of competing methods to sub-optimal
values, or choose a variant of the method that is known by experts to
be sub-optimal. They may also select competing methods in a
sub-optimal way, i.e. consciously or sub-consciously exclude the best
methods from the comparison. Beyond the problems of technical
expertise and optimization bias, interpretation and representation
issues might also affect the final conclusions of a comparison
study. Given the same quantitative outputs, the impression of
the reader can be affected, e.g., by the choice of the vocabulary in
the results section, by graphical representation, or by the choice of
the main quantitative criterion used to compare the methods.

For all these reasons, most comparison studies published in the
literature as part of an original paper are substantially
biased. These problems stress the importance of \lq\lq neutral
comparison studies'' that we define as follows:
\begin{itemize}
\item[A.] The main focus of the article is the comparison itself. It
  implies that the primary goal of the article is not to introduce a
  new promising method.
\item[B.] The authors should be reasonably \lq\lq neutral''. For example,
  an author who has published an article on a new method six months
  before is likely to be less neutral than an author who has often
  used several of the considered methods for statistical consulting
  and, say, previously investigated three of them more precisely in
  methodological articles.
\item[C.] The evaluation criteria, methods, and data sets should be chosen in a rational way, see Section 4 for a more extensive discussion of this problem.
\end{itemize}

Note that the comparison between the competing methods is essentially
not affected by the bias discussed in the introduction. Hence,
one idea could be to extract neutral comparisons from
comparison studies included in original articles presenting new
methods---by considering the competing methods only. However, one
should keep in mind that these methods probably have not been given as
much attention as in the case of a real neutral comparison study that
does not involve any new method. This relative lack of attention
possibly leads the underestimation of their performance.

To come to the point, in an original article on a new method,
the focus is on the new method, and that is where the authors
generally spend most of their energy. Consequently, comparisons
between competing methods should not be over-interpreted because they
may be of sub-optimal quality. On this account we make a (passionate)
plea for neutral comparison studies in computational sciences.


\section{Tidy neutral comparison studies}

In the same way clinical research and clinical studies have to
  be well planned and executed (following strict guidelines), comparison
  studies should also follow a well-defined study design. They should
  be based on a sound theoretical framework, appropriate analysis
  methods, and carefully selected components. There is a variety of
  literature on the design and analysis of comparison studies
  available---we propagate, for example,
  \citet{Hothorn+Leisch+Zeileis+Hornik@2005} as a theoretical
  framework and \citet{Eugster+Hothorn+Leisch@2012} as its practical
  implementation. However, regardless of the concrete framework,
  general considerations on the individual components---evaluation
  criteria, methods and method parameters, and data sets---can be
  made.
  
\begin{itemize}
  \item \textit{Choice of evaluation criteria:} In the case of supervised learning algorithms, simple evaluation
    criteria are, e.g., the error rate or preferably the area under curve that is based on the predicted class probabilities. Such criteria are natural and
    objective. However, many other criteria have an impact on the
    usefulness of a method in practice for applications. From a
    pedagogical point of view, one should not forget that the method
    is destined to be used by experts or non-expert
    users. Therefore, all other things being equal, simplicity of a method constitutes an important advantage, similarly to the clinical context where the simplicity of a therapy protocol should be seen as a major advantage.  From a
    technical point of view, particular attention may be devoted to
    computational aspects such as computation time and storage
    requirements (similarly to the costs in the clinical context), the influence on initial values in an iterative
    algorithm, or more generally the dependence on a random
    generator (similarly to the robustness of the therapy's effect against e.g. technical problems or human errors). 

  \item \textit{Choice of methods and method parameters:} 
    The choice of methods is a very subjective one.  At any rate,
      the concrete choice should be clearly motivated and personal
      preferences and similar influences should be clearly acknowledged. 
      Researchers are inevitably conducted by personal preferences, past 			experiences and own technical competence. 
      However, the choice should also be guided by objective arguments. Possible criteria are i) the popularity of the methods in practice (for instance: restrict to methods that have been used in at least three concrete studies), 
ii) results available from the literature (e.g. from a previous comparison study) to pre-filter good candidates, or iii) specific pre-defined criteria specifying the nature of the method, for example \lq\lq only statistical regression-based methods''.  
None of these criteria should be considered as mandatory for a neutral comparison study. But we claim that, the set of criteria being defined, the methods should be more or less randomly sampled within the range of available methods.
      
      As far as method parameters like hyperparameters are concerned, they should   be chosen based on ``standard practice rules''.

  \item \textit{Choice of data sets:} Researchers performing
    comparison studies also choose data sets. Considering the high
    variability of relative performance across data sets, a comparison
    study based on different data sets (all other things being equal)
    may obviously yield different results. Variability arises both
    because error estimation with standard resampling-based estimators
    is highly variable for a given underlying joint distribution of
    predictors and response \citep{dougherty2010performance} and
    because different data sets also have different underlying
    distributions.
      Therefore, it is important to make an ``as representative as
      possible'' selection of data sets to cover the domain of
      interest. At best, the data sets are chosen from a set of data
      sets representing the domain of interest using standard sampling
      methodology.
    
\end{itemize}

In summary, many choices have to be met when performing a comparison
study, for example, in the case of supervised learning with
high-dimensional data: the included methods (e.g. penalized
regression, tree ensembles, support vector machines, partial least
squares  dimension reduction, etc), the considered variants (which
kernel for SVM, which fitting algorithm for penalized regression,
which optimality criterion for PLS, which splitting criterion for tree
ensembles, etc), the data domain (which type of data sets), the
parameter tuning procedure (which resampling scheme, which candidate
values). With this in mind, it is clear that the topic of interest
cannot be handled completely by a single comparison study. Different
comparison studies with similar scope may yield different
conclusions. This can be seen as a limitation of each single
comparison study -- or as an argument to perform more such comparison
studies. Going one step further in the comparison with clinical
research, one could also imagine a concept of meta-analysis for
comparison studies in computational sciences. In the clinical context,
meta-analyses provide a synthesis over different populations,
different variants of the investigated therapies, different technical
conditions, different medical teams, etc. Similarly, meta-analyses for
computational studies in computational sciences would provide
syntheses over different data domains, different variants of the
considered methods, different software environments, different teams
with their own areas of expertise, etc.

\section{Negative results}
Comparison studies can be a good vehicle for negative research findings. Publication biases and the necessity
to ''accentuate the negative'' \citep{ross2009trial} are
well-documented in the context of medical and pharmaceutical
research. In applied statistics and data analysis research, however,
this issue receives very poor attention \citep{boulesteix2010over},
even if the publication of negative results may be extremely useful
in many cases.

The systematic exclusion of negative results from publication might in
some cases be misleading. For example, imagine that ten teams around
the world working on the same specific research question have a
similar promising idea that in fact does not work properly for any
reason. Eight of the ten teams obtain disappointing results. The ninth
team sees a false positive in the sense that they observe significant
superiority of the new promising method over existing approaches
although it is in fact not better. The tenth team optimizes the
method's characteristics \cite{jelizarow2010over}
and thus also observes significant superiority. The two latter teams
report the superiority of the promising idea in their papers, while
the eight other studies with negative results remain unpublished: a
typical case of publication bias. This scenario is certainly
caricatural, but similar things are likely to happen in practice
although in a milder form. Note that it is very difficult to give
concrete examples at this stage, since such stories essentially remain
unpublished.

Nevertheless, the publication of negative results might entail
substantial problems. Most researchers (including ourselves!) probably
have more ideas that turn out to be disappointing than ideas that work
fine. Try-and-error is an essential component of research. It would
thus be impossible (and uninteresting anyway) to publish all negative
results. But then, what was promising and what was not promising? What
is likely to interest readers and what was just a bad idea that nobody
else would have thought of? Obviously this decision that would have to
be taken by reviewers and editors is a subjective one. Assessing
whether a new method with negative results deserves publication in a
separate paper is anything but trivial. With this in mind, we believe
that the publication of negative findings within large well-designed
comparison studies would be a sensible compromise in order to diffuse
negative findings without congesting the literature with negative
papers.

Journals would not have to fear for their impact, since good comparison
studies are usually highly accessed and cited. Authors would not be urged to
make something out of their promising idea on which they have spent a
lot of time: a large comparison study would be an alternative to
publish important results and share their vast experience on the topic
without fishing for significance. And "fishing for significance" would
lose part of its attractiveness. Most importantly, readers would be
informed about important research activities they would not have heard
of otherwise.

Note that ``standard practice rules'' in computational sciences (e.g., regarding the choice of method parameters) are often implicitly the result of comparison studies. For instance, a \lq\lq standard parameter value'' becomes standard because it yields better results than another value. In other words, negative results are often hidden behind standard practice rules - most of them
remaining unpublished. Our point is that this process could be made more transparent and more informative for the readers if these negative results were published within extensive comparison studies.

Drawing the comparison with clinical research from the introduction even further, we also think that it may be interesting to publish articles on \lq\lq pitfalls''. By \lq\lq pitfall'' we mean the   inconveniences of a data analysis method such as, e.g., a non-negligible bias, a particularly high variability, or
  non-convergence of an algorithm in specific cases that may lead to
  misleading results. By \lq\lq negative result'' we mean a
  disappointing result of a new method that had been considered as
  promising problem solver for a specific case.
In computational literature such research results are often hidden in
the middle of articles that are actually devoted to something
else. This is in contrast to clinical research, where pitfalls of
existing methods (e.g. an adverse effect of a drug) may be the main
object of an article, even if no alternative solution  is proposed
(for example in form of an alternative drug).

\section{Limitations}

Neutral comparison studies are in our opinion crucial to make the
establishment of standards more objective and to give a chance to
methods that are at first view unspectacular and would otherwise be
pigeonholed. However, comparison studies and their impact should not
be over-interpreted. Firstly, one should not forget that no method is
expected to work well with all data sets (the well-known ``no
  free lunch theorem''). Hence, a method that scores well in many
comparison studies may do poorly in a specific data set. Comparison
studies are not expected to yield an absolute truth applicable to all
situations. They are solely useful to determine general trends that
may be useful to the community to select a set of standard methods
that often perform well.

Secondly, comparison studies are essentially limited because they rely
on the specific and sometimes arbitrary choices regarding the study
design: the choice of simplifying evaluation criteria that probably do
not reflect the complexity of concrete data analysis situations, the
choice of method parameters that may substantially impact the relative
performance of the considered methods, and last but not least the
choice of specific example data sets.

Thirdly, comparison studies are often underpowered in the sense that the number of included data sets is insufficient considering the high variability of performance across data sets. With a few exceptions (see \cite{desouza2010comprehensive} for a comparison of machine learning algorithms based on 65 gene expression data sets), comparison studies often include up to 10-15 data sets, which is probably not enough.  This issue may be further investigated in future research.

Fourthly, comparison studies essentially ignore the substantive context
of the data sets they consider. Data sets are sometimes preprocessed
without much knowledge of the signification of the variables. All
methods are applied in the same standardized way to all data sets. The
analysis is thus intentionally over-simplified. An important aspect of
the data analysis approach is neglected, which does not reflect the
complexity and subtleties of the data analyst's work
\citep{keiding2010reproducible}. A method that does not work well if
applied in a standard way without knowledge of the substantive context
might perform better in concrete situations, hence reducing the
relevance of comparison studies.

\section{Conclusion}

Neutral comparison studies are often considered as less exciting than
project on new methods by both researchers and journal editors -- but
not by readers. They can neither be expected to always give the best
answer to the question \lq\lq which method should I use to analyze my
data set'' nor reflect a real data analysis approach that takes the
substantive context into account.  However, we believe that they may play a crucial role to make the evaluation of existing methods more rational and to establish
standards on a scientific basis. They certainly deserve more
consideration than is currently the case in the literature.

\bibliography{letter_compstud}

\end{document}